# Superconducting nanowire single photon detector with on-chip bandpass filter

Xiaoyan Yang, Hao Li, Weijun Zhang, Lixing You*, Lu Zhang, Xiaoyu Liu, Zhen Wang, Wei Peng, Xiaoming Xie, and Mianheng Jiang

*State Key Laboratory of Functional Materials for Informatics, Shanghai Institute of Microsystem and Information Technology (SIMIT), Chinese Academy of Sciences, 865 Changning Rd., Shanghai 200050, China*
*[*lxyou@mail.sim.ac.cn](mailto:lxyou@mail.sim.ac.cn)*

**Abstract**: Dark count rate is one of the key parameters limiting the performance of the superconducting nanowire single photon detector (SNSPD). We have designed a multi-layer film bandpass filter that can be integrated onto the SNSPD to suppress the dark counts contributed by the stray light and blackbody radiation of the fiber. The bandpass filter is composed of 16 $SiO_2$/Si bilayers deposited onto the backside of a thermally oxidized Si substrate. The substrate shows an excellent bandpass filter effect and provides a high transmittance of 88% at the central wavelength of the pass band, which is the same as that of the bare substrate. The SNSPDs fabricated on the substrate integrated with the bandpass filter show conspicuous wavelength-sensitive detection efficiency. The background dark count rate is reduced by two orders of magnitude to sub-Hz compared with the conventional SNSPD (a few tens of Hz). The detector exhibits a system detection efficiency of 56% at DCR of 1 Hz, with the measured minimal noise equivalent power reaching $2.0 \times 10^{-19}$ W/Hz$^{1/2}$.

## 1. Introduction

High-performance superconducting nanowire single photon detectors (SNSPDs) have been widely demonstrated as high-performance single-photon detector (SPD) in many fields such as quantum information, quantum optics, free space laser communication and light detection and ranging [1-6]. The false detection events or dark count rate (DCR), which represents the noise level of the SNSPD, is one of the key parameters for practical applications. For example, the quantum bit error rate (QBER) in quantum key distribution (QKD) is proportional to DCR. A low DCR may further extend the communication distance of QKD while keeping a specific low QBER [7].

In the past two years, the detection efficiency (DE) of SNSPD has been greatly improved owing to the significant developments in its structure and material innovation [8-11]. However, the reduction in DCR still lacks improvement. As a matter of fact, DCR has become higher with the increase of DE. Up to now, the typical background DCR of SNSPD ranges from a few tens of Hz to several hundred Hz.

The DCR of SNSPD is usually composed of two parts, namely, the intrinsic DCR (iDCR) and extrinsic DCR (eDCR). The iDCR is related to the spontaneous vortex motion in the nanowire [12, 13] and is exponential to the normalized bias current. On the other hand, the eDCR is caused by the noise photons from both the blackbody radiation of the fiber itself and stray light penetrated into the fiber. The iDCR can be suppressed by decreasing either the operation temperature or the bias current [14]. Therefore, the eDCR usually plays a key role, under such a condition where iDCR is negligible. Even through the eDCR originating from the stray light can be removed by using armored fiber, the eDCR contributed by the blackbody radiation of the fiber is hardly avoided as long as the fiber is attached to the SNSPD. Shibata et al. effectively suppressed eDCR by using a cold bulk optical filter. However, this compromised the system detection efficiency (SDE) by a loss of 2.7 dB [15].

In this paper, we present a novel method for suppressing the eDCR by introducing a multi-layer film bandpass filter onto the SNSPD devices. The eDCR caused by blackbody radiation of fiber can be effectively suppressed by two orders of magnitude to less than 1 Hz, while still maintaining the high SDE of the SNSPD. The properties of the filter and SNSPD are discussed in detail.

## 2. Description of the bandpass filter and its characteristics

The current state-of-the-art optical film technology offers us the possibility of designing and fabricating different kinds of optical filters. Since the eDCR is caused by the thermally induced photons of a broad spectrum, the use of a narrow bandpass filter should let the signal photons at a specific waveband pass with little attenuation, while blocking the noise photons of all the other wavebands. Because the peak intensity of the blackbody radiation from the fiber at 300K is longer than 2 μm, the stopband of an ideal bandpass filter should cover a wavelength range from a few hundred nanometers up to several microns. However, a narrow bandpass filter with a wider stopband is more difficult to design and fabricate than that with a narrower stopband. Considering the wavelength-dependent detection efficiency [16] and a reasonable filter complexity, we designed and fabricated a bandpass filter in the wavelength range of 500–2200 nm with the pass band of 1545-1555 nm.

The design principle of the bandpass filter is the interference effect in dielectric multi-layer films [17]. The thicknesses of the Si ($SiO_2$) layers usually range from zero to one wavelength, which corresponds a light wave phase change ranging from 0 to $2\pi$. Using the optical film design software Filmstar[tm], the transmittance was calculated based on the transfer-matrix method and the genetic algorithm was performed to optimize the precise thickness of each layer [17-19]. The practical bandpass filter is composed of periodic $SiO_2$/Si bilayers alternatively deposited layer-by-layer on the backside of a thermal oxidized Si substrate, which is a popular substrate for the fabrication of SNSPD [9]. The schematic view of the bandpass filter fabricated on the substrate is shown as Fig. 1. The filter is composed of 32 layers, providing a total thickness of 6.02 μm. The thickness of the Si layers varies from 54 nm to 432 nm, while the thickness of the $SiO_2$ layers varies from 137 nm to 308 nm. The calculated transmittance of the substrate with the bandpass filter is shown as the blue curves in Fig. 2. The passband (1545–1555 nm) transmittance of the substrate with the filter on the backside (from air to air) is calculated to be 93% and the full width at half maximum value is around 30 nm. The average transmittance is lower than 0.1% for the stopbands in the ranges of 500–1500 and 1600–2200 nm. The performance of the substrate with the filter was measured at room temperature using spectrophotometer (PE Lambda950). The corresponding results are shown in Fig. 2, indicated by the black curves. The experimental results show a peak transmittance of 88% at the wavelength of 1545 nm. The discrepancy of the filter characteristics between the calculation and the measurement may be explained by the imperfect thickness control and the refractive index variance in the multi-layer film sputtering. Though the measured peak transmittance is 5% lower than the calculated value, it reaches the same transmittance of the substrate without filter which is shown as the red curves in Fig. 2.

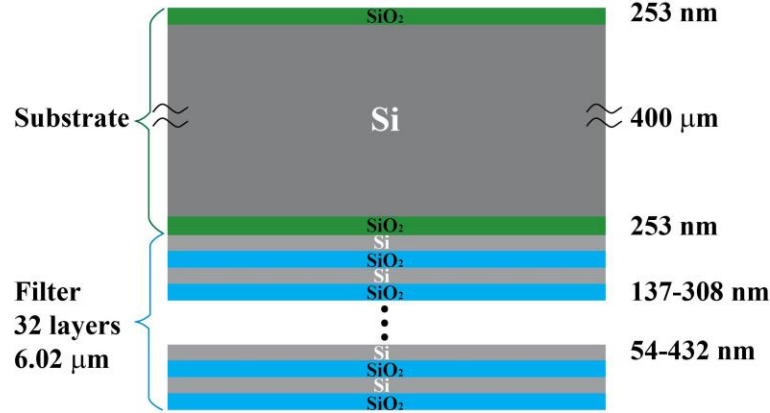

Fig. 1. Schematic structure of the bandpass filter on the backside of a thermal oxidized Si substrate.

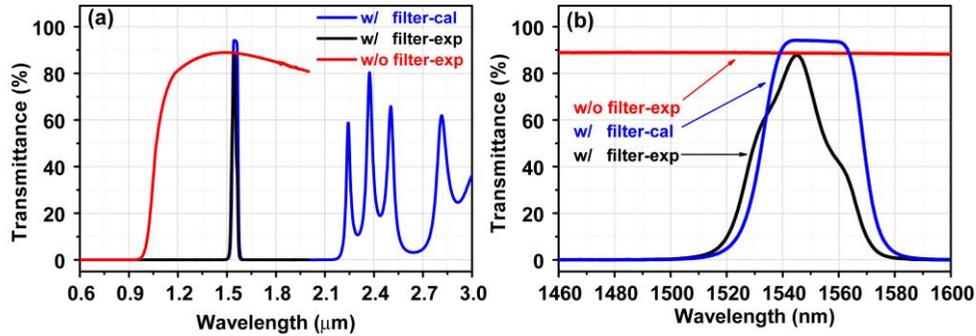

Fig. 2. Calculated and experimental optical characteristics of the substrate with and without the bandpass filter at different wavelength scales. (a) 0.6–3 μm; (b) 1460–1600 nm. The simulation is based on the parameters at the room temperature. Limited by the instrument, the measurement was carried out for the wavelength from 0.5 μm to 2.0 μm at room temperature.

## 3. SNSPD with the bandpass filter

In order to test the effect of the bandpass filter on the performance of the SNSPD, we fabricated SNSPDs on thermal oxidized Si substrates with and without the bandpass filter in the same batch. A 6.5 nm-thick superconducting NbN film was chosen as the detector material. The sensitive area has a conventional meandered nanowire structure with linewidth/space of 90/110 nm and size of 15 × 15 μm. A lensed optical fiber was vertically aligned to the sensitive area from the backside of the substrate through the filter. The distance between the fiber tip and the substrate is controlled to be 20 ± 5 um at the room temperature. The measurement was performed at the temperature of 2.3 K using a cryostat based on a compacted Gifford–McMahon cryocooler. During the DCR measurement, the stray light was shielded with armored fiber. This arrangement ensures that the eDCR represents the dark counts contributed by the blackbody radiation of the fiber.

A tunable continuous laser (Agilent 81980A) is adopted for studying the wavelength dependence of the SNSPD. The wavelength of the laser can be tuned with a step of 1 nm from 1465 nm to 1575 nm. To measure the SDE, the power of the laser is heavily attenuated to 1 M photons ($N_{ph}$) per second. The response signal of SNSPD is amplified then fed to a photon counter (Stanford SRS400) for counting. The SDE is determined by the difference of the counted photon number and DCR divided by $N_{ph}$. Figure 3 shows the wavelength dependence of SDE for the two SNSPDs (#A and #B) with and without a bandpass filter, respectively. The data was recorded at a dark count rate of 100 Hz for each detector. For the device #B without the bandpass filter, the SDE is at around 50% for the measured waveband. The slight oscillation of SDE could be attributed to the light interference of the substrate. However, the device #A showed prominent bandpass filter effect, with a profile similar to the transmittance data shown in Fig. 2(b). The wavelength-dependence of the SDE proves the proper functioning of the bandpass filter. It could be noted that the wavelength of the maximal SDE is shifted from 1545 nm to 1532 nm, owing to the temperature influence of the filter. In principle, the refractive indices of Si and $SiO_2$ decrease with temperature [20, 21]. Optical simulation studies indicate that the change in refractive index will induce wavelength shift in the passband of filter. However, the transmittance of the passband will remain unchanged.

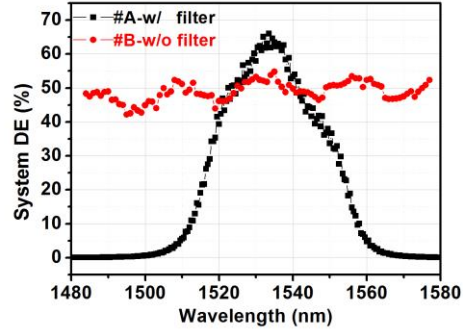

Fig. 3. Wavelength dependence of SDE in SNSPDs with and without the bandpass filter at DCR of 100 Hz.

Figure 4(a) shows the current dependence of the SDE, iDCR and DCR in the detectors #A and #B. Both the detectors show maximal SDE of around 60%, which indicates the efficiency of the fabrication process and absence of negative influence caused by the filter. iDCR and DCR follow each other at high bias current and deviation happens at lower bias current due to the contribution of eDCR in DCR for both detectors. More interestingly, the inflection points where the deviation happens for the two detectors are different. Owing to the bandpass filter effect, the maximal DCR of the detector #A caused by the blackbody radiation of fiber is effectively suppressed to 0.4 Hz. On the other hand, in the detector #B without the bandpass filter, the maximal DCR caused by the blackbody radiation has a value of around 20 Hz. These results indicate that the eDCR caused by the blackbody radiation is suppressed by two orders of magnitude with the use of the bandpass filter. To further establish a straightforward comparison of the performance of both detectors from the viewpoints of application, we analyzed the relation of SDE as a function of DCR, as shown in Fig. 4(b). At DCR of 100 Hz, the detectors #A and #B have comparable SDE values of 64% and 50%, respectively. However, when DCR is 1 Hz, there is an obvious difference in the SDEs of the detectors #A and #B, showing a value of 56% and 6%, respectively. Table 1 summarizes the SDEs of the two detectors at different DCR. As clearly evidenced from the results, the use of the on-chip bandpass filter effectively suppresses the eDCR to sub-Hz while still maintaining a high SDE. This obviously satisfies the requirements of most of the applications, which demand SPDs with both high SDE and low DCR.

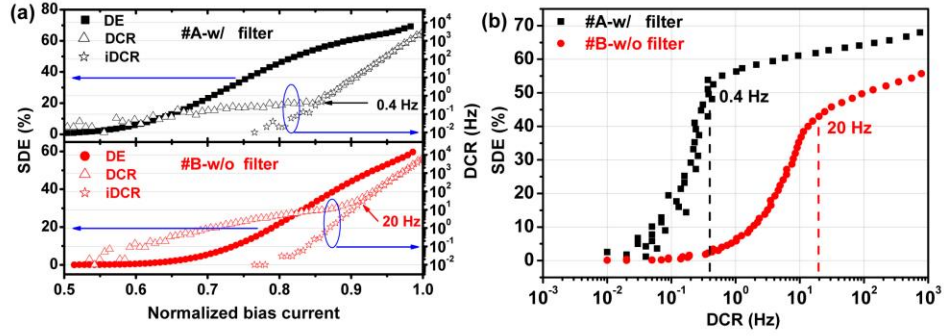

Fig. 4. (a) Normalized bias current dependence of SDE, iDCR and DCR for detectors #A and #B; (b) SDE vs DCR for detectors #A and #B.

**Table 1. Performance of detectors #A and #B**

| DCR (Hz) | | 0.1 | 0.4 | 1 | 10 | 100 |
|---|---|---|---|---|---|---|
| SDE (%) | #A | 16.8 | 52.4 | 56.2 | 61.2 | 64.2 |
| | #B | 0.7 | 2.6 | 6.0 | 36.4 | 50.2 |
| NEP (W/Hz$^{1/2}$) | #A | $3.7 \times 10^{-19}$ | $\mathbf{2.0 \times 10^{-19}}$ | $3.0 \times 10^{-19}$ | $8.8 \times 10^{-19}$ | $2.6 \times 10^{-18}$ |
| | #B | $6.7 \times 10^{-18}$ | $4.4 \times 10^{-18}$ | $2.8 \times 10^{-18}$ | $\mathbf{1.6 \times 10^{-18}}$ | $3.6 \times 10^{-18}$ |

In general, noise equivalent power (NEP) is an important parameter in evaluating the performance of an ultrasensitive detector. The NEP of SNSPD is defined to reveal the interplay between the dark count rate and the detection efficiency [22], and is expressed as $NEP = h\nu \cdot \sqrt{2DCR} / DE$, where $h$, $\nu$ represent the Plank constant, and frequency of the photon, respectively. The NEP of the detectors #A and #B at different DCR is also shown in Table 1. As indicated in the table, the lowest measured NEP reached $2.0 \times 10^{-19}$ W/Hz$^{1/2}$ at DCR of 0.4 Hz for the detector #A with SDE of 52%. This is among the lowest system values reported for meander-structured SNSPD devices [9-11].

It is noted that the dependence of SDE on the bias current for detectors A and B are not identical in overall shape though all the fabrication parameters are same and the SEM images gave the same linewidth of the nanowires. Actually, the SDE performance varies even for the different SNSPDs on the same wafer. A possible explanation is the local nonuniformity of the film for one detector to the other. It is still necessary to further improve the quality of the film.

## 4. Conclusion and outlook

In summary, we have designed and fabricated an on-chip optical bandpass filter on the backside of the SNSPD to suppress the dark counts caused by the blackbody radiation of the fiber. The dark counts are effectively blocked so that the background DCR is suppressed from a few tens of Hz to sub-Hz. The detector with the bandpass filter has SDE of 56% at DCR of 1 Hz, with the measured NEP reaching a record low value of $2.0 \times 10^{-19}$ W/Hz$^{1/2}$ at DCR of 0.4 Hz. Indeed, the on-chip bandpass filter is also effective for the free-space optical coupled SNSPD since the ambient blackbody radiation may be more serious than that in fiber-coupled SNSPD. Future work is underway to optimize the design of the bandpass filter with an aim to increase the practical bandwidth of the filter, taking into consideration the temperature influence of the filter during the design.

**Acknowledgments**

We acknowledge the technical support provided by F. W. Gan of SIMIT and S. J. Zhang of East China Normal University. This work is funded by the National Natural Science Foundation of